\documentclass[aps,prl,showpacs,reprint,groupedaddress]{revtex4-1}

\usepackage{graphicx}
\usepackage{amsmath}
\usepackage{natbib}
\usepackage{amssymb}
\newcommand{\degp}{\ensuremath{{\cal P}}}
\newcommand{\pardef}{\ensuremath{\stackrel{\bigtriangleup}{=}}}
\begin{document}

\newcommand*{\mean}[1]{\left\langle #1 \right\rangle}
\newcommand*{\ket}[1]{\left|#1\right\rangle}
\newcommand*{\bra}[1]{\left\langle #1\right|}
\newcommand*{\abs}[1]{\left|#1\right|}
\newcommand*{\tr}{\mathrm{tr}}
\newcommand*{\diag}{\mathrm{diag}}
\newcommand*{\gr}[1]{\mathbf{#1}}
\newcommand*{\grm}[1]{\mathrm{\mathbf{#1}}}
\newcommand*{\norm}[1]{\left\Vert #1 \right\Vert }
\newcommand*{\DPZ}[3]{\left.\frac{\partial#1}{\partial#2}\right|_{#3}}
\renewcommand*{\i}{\mathrm{i}}


\title{Depolarization remote sensing by orthogonality breaking}

\author{Julien Fade}
\email{julien.fade@univ-rennes1.fr\\}
\affiliation{Institut de Physique de Rennes, Universit\'{e} de Rennes 1, CNRS, Campus de Beaulieu, 35042 Rennes, France}
\author{Mehdi Alouini}
\affiliation{Institut de Physique de Rennes, Universit\'{e} de Rennes 1, CNRS, Campus de Beaulieu, 35042 Rennes, France}
\date{\today}

\date{\today}

\begin{abstract} 
  A new concept devoted to sensing the depolarization strength of
  materials from a single measurement is proposed and successfully
  validated on a variety of samples. It relies on the measurement of
  the orthogonality breaking between two orthogonal states of
  polarization after interaction with the material to be
  characterized. The two fields orthogonality being preserved after
  propagation in birefringent media, this concept is shown to be
  perfectly suited to depolarization remote sensing through fibers,
  opening the way to real time depolarization endoscopy.

\end{abstract}

\pacs{42.25.Ja; 42.62.-b; 07.60.Vg; 07.07.Df}

\maketitle

Since early times, the interaction of the electromagnetic radiation
with natural objects has been the subject of intensive
investigations. Most significant advances in terms of physics
understanding and applications are in the microwave domain where the
amplitude, the phase and the polarization of the field are routinely
used, for instance, in long range sensing systems in order to optimize
the extracted information from a given scene
\cite{hal80,dal91,cou94,pla11}. Conversely, in the optical domain
where common sensors are based on quadratic detection, it is usual to
cope with the intensity of the backscattered field. In particular in
the case of polarimetric imaging, the determination of the state of
polarization of the backscattered light involves the Stokes vector
whose four elements correspond to measured intensities \cite{sol81}. A
rigorous analysis of the depolarization nature of a given material
implies the determination of its Mueller matrix which linearly couples
the backscattered Stokes vector to the illumination Stokes vector,
thus requiring 16 measurements \cite{pez95}. Consequently, this common
approach is very stringent since it implies a perfect control of the
polarization states of the emitted light as well as a precise
projection of the backscattered electric field on the four analysis
states of polarization. Moreover, this approach is very restrictive in
terms of wavelength tuning and applies only to free space propagation
forbidding the use of fibers as part of the polarimetric apparatus.

Although some techniques have been recently proposed to perform
polarimetric measurements through fibers \cite{des09,oh08}, they are
again based on the same principle, namely, shining the object with a
diversity of states of polarization and analyzing the backscattered
signal with a polarization sensitive detector. In Ref. \cite{des09},
it is demonstrated that the degree of polarization of a given material
can be recovered statistically but to the expense of a large and time
consuming number of realizations. In this context, we felt that, as
far as the depolarization strength (degree of polarization) of a
material is the parameter of interest, the usual approaches are not
optimal from an experimental point of view. Indeed, as far as one
parameter has to be determined, it should be possible in principle to
retrieve it from a single measurement. More importantly, if such a
single measurement is achievable experimentally, then it automatically
solves the problem of polarimetric remote sensing through fibers. 
In this Letter we propose to revisit the way of performing
depolarization measurements in the optical domain using the optical
electric field rather than its intensity. To this aim, we propose a
novel polarimetric sensing modality which involves the concept of
polarization orthogonality breaking. We show theoretically and confirm
experimentally that this new sensing modality is able to provide the
depolarization nature of a material and is by essence insensitive to
propagation through fibers, opening the way to real time endoscopic
polarimetric imaging.

In a majority of applications, polarimetric imaging end-users are
mainly interested in revealing polarization contrasts which may not
appear on standard intensity images \cite{sol81,gro99,bre99}. In such
context, characterizing the full Mueller matrix or the Stokes vector
$\grm{S}$ provides superfluous information, as evidenced by the
variety of simplified polarimetric imaging designs available
\cite{gro99,bre99,jac02,fad12a}. In most applications, a relevant
contrast parameter to consider is the degree of polarization (DOP)
$\degp=\bigl[1-4\det( \Gamma_{\mathrm{out}})/\tr(\Gamma_{\mathrm{out}})^2\bigr]^{1/2}$,
with $\Gamma_{\mathrm{out}}=\langle \grm{E}_{\mathrm{out}}\grm{E}_{\mathrm{out}}^\dagger\rangle $
denoting the polarization matrix of the backscattered light which can
be bijectively derived from $\grm{S}$ \cite{sol81}, with $\langle
\rangle$ denoting statistical (ensemble) average. However, a majority
of materials can be considered as purely depolarizing
\cite{bre99,lu96}, i.e., their Mueller matrix is well approximated by
a diagonal matrix of rank 2. In that case, $\degp$ can be evaluated
from only two intensity measurements through orthogonal polarizers
$\degp=(I_{//}-I_{\perp})/(I_{//}+I_{\perp})$, when the samples are
enlightened with linearly polarized light. In general, the action of a
purely depolarizing material on an incident state of polarization
$|\grm{E}_\mathrm{in}\rangle$ can be phenomenologically modeled by a partial
projection of the incident state onto the orthogonal polarization
direction denoted $|\grm{E}^\perp_{\mathrm{in}}\rangle$, i.e.,
$|\grm{E}_{\mathrm{out}}\rangle = \sqrt{\rho} (|\grm{E}_\mathrm{in}\rangle +
\alpha|\grm{E}^\perp_\mathrm{in}\rangle)$, where
$\alpha=\sqrt{1-\degp}/\sqrt{1+\degp}$. Using Jones matrix formalism,
such relation reads
\begin{equation}\label{Jones}
|\grm{E}_{\mathrm{out}} \rangle=\grm{J_m} |\grm{E}_{\mathrm{in}} \rangle=\sqrt{\rho}\bigl[\grm{R}_{|\grm{E}_{\mathrm{in}}\rangle}\bigr]^\dagger \begin{pmatrix}1&\alpha\\ \alpha & 1\end{pmatrix}\grm{R}_{|\grm{E}_{\mathrm{in}}\rangle}|\grm{E}_{\mathrm{in}} \rangle,
\end{equation} 
where the unitary matrix $\grm{R}_{|\grm{E}_\mathrm{in}\rangle} \in SU(2)$
corresponds to the generalized polarization rotation mapping the
transverse field eigen basis $\{|\grm{e}_1\rangle,|\grm{e}_2\rangle\}\pardef\{|\grm{e}_{\rm in}\rangle,|\grm{e}^\perp_{\rm in}\rangle\}$ into the standard linear transverse field basis $\{|\grm{e}_X\rangle,|\grm{e}_Y\rangle\}$.

\begin{figure}[htbp]
\includegraphics[width=7cm]{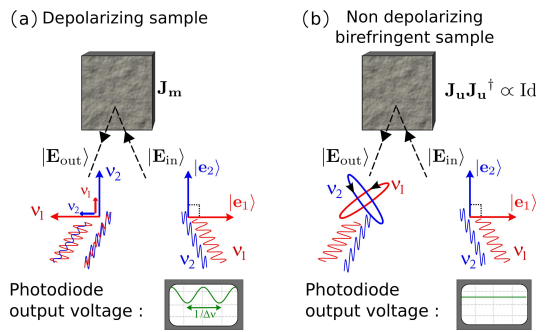}
\caption{Principle of Depolarization Sensing by Orthogonality Breaking
  (DSOB).}
\label{fig1}
\end{figure}

From Eq.(\ref{Jones}), it is seen that a non-null off-diagonal term
$\alpha$ leads to a superposition of the two orthogonal states of the
electromagnetic field. Assuming that these two eigenstates
$|\grm{e}_1\rangle$ and $|\grm{e}_2\rangle$ are non-degenerated, that
is, they have two distinct eigenfrequencies, a coherence oscillation
at the eigenfrequencies difference is expected, as illustrated in
Fig.\ref{fig1}.(a). Moreover, the amplitude of the coherence
oscillation is directly linked to the depolarization value $\degp$
through $\alpha$. Thus, the depolarization strength of a given
material can be recovered from the way this material breaks the
orthogonality of a properly prepared state of the electromagnetic
field.  To illustrate this property, let us consider a light source
emitting two distinct frequencies along two orthogonal polarization
states
\begin{equation}\label{Ein}
  |\gr{E_{\mathrm{in}}}(\boldsymbol{r},t) \rangle= \frac{E_0} {\sqrt{2}} \Bigl[ \psi_1(\boldsymbol{r},t) |\grm{e}_1 \rangle  + \psi_2(\boldsymbol{r},t) |\grm{e}_2 \rangle  \Bigr],
\end{equation}
with $\psi_k(\boldsymbol{r},t)= M_k(\boldsymbol{r}) e^{-2j\pi \nu_k
  t}$. Such state is a physical implementation of a non-quantum
entangled state \cite{sim10,qia11}. For the sake of simplicity, we consider
the case of orthogonal states of equal intensities, which can be shown
to be the best compromise in terms of detection performance, and we
restrict ourselves to the case of plane waves, i.e.,
$M_k(\boldsymbol{r})=1$. In the most general case, the eigenstates
$|\grm{e}_1\rangle$ and $|\grm{e}_2\rangle$ can be elliptical with
orthogonal azimuth directions and equal ellipticity $\phi$.  The
polarization matrix $\Gamma_{\mathrm{in}}$ of such illumination field can be
decomposed in two terms $\Gamma_{\mathrm{in}}=\Gamma^0+\Gamma^{\Delta \nu}
e^{-j2\pi \Delta \nu t}$. The first term $\Gamma^0$ accounts for the
continuous part of second-order correlations between the field
transverse components, whereas $\Gamma^{\Delta \nu}$ describes
interference terms, oscillating at the beat note frequency $\Delta \nu
= \nu_2-\nu_1$. From the above definitions, one can easily derive the
input field continuous intensity $I^0_{\mathrm{in}}=\tr (\Gamma^0)=|E_0|^2$,
while the amplitude of the beat note intensity modulation
$I_{\mathrm{in}}^{\Delta\nu}=\tr(\Gamma^{\Delta\nu})=0$, since the two
orthogonal input states do not interfere.

When such a field interacts with an object characterized by its Jones
matrix $\mathrm{\mathbf{J}}$, the value of the beat note intensity
modulation of the resulting field
$I_\mathrm{out}^{\Delta\nu}=\tr(\grm{J}\Gamma^{\Delta\nu}
\grm{J}^\dagger)$ provides a direct information on the depolarization
properties of the object. As illustrated in Fig.\ref{fig1}.(a), the
loss of orthogonality induced by the interaction with a depolarizing
sample gives rise to an interference intensity component oscillating
at a frequency $\Delta \nu$ and whose amplitude is bijectively related
to the material depolarization strength. If the depolarizing sample is
described by $\grm{J_m}$ given in Eq.(\ref{Jones}), one indeed has
\begin{equation}\label{defC}
  C_\mathrm{out}^{\Delta\nu/0} \pardef \frac{P_\mathrm{out}^{\Delta\nu}}{P_\mathrm{out}^{0}}=\frac{4\alpha^2}{(1+\alpha^2)^2}\cos^2 \phi=(1-\degp^2)\cos^2 \phi.
\end{equation}
The DOP of the sample can thus be directly retrieved from a
single measurement of the beat note optical power
$P_\mathrm{out}^{\Delta \nu}$ (normalized by the cw optical power
$P_\mathrm{out}^{0}$). This quantity can be measured with a heterodyne
detection, thus enabling fast ($<1 \ \mu$s for $\Delta \nu = 1$ GHz),
highly-sensitive depolarization measurements.

On the other hand, non-depolarizing media (isotropic, birefringent or
optically active media) are characterized by unitary trace-preserving
Jones matrices. Such a matrix verifies $\grm{J_u}\grm{J_u}^\dagger
\varpropto \mathrm{Id}$, where $\mathrm{Id}$ is the identity matrix,
and thus $I_\mathrm{out}^{\Delta\nu}=0$, demonstrating that
orthogonality between the two illumination states is preserved when
light propagates through such media. This property, illustrated in
Fig.\ref{fig1}.(b), is valid provided no significant dispersion appears
between frequencies $\nu_1$ and $\nu_2$. This is very unlikely to
occur in practice since $\Delta \nu$ will not exceed tens of GHz for
the beat note to be detectable on a photodetector. A consequence of
this result is that the Depolarization Sensing by Orthogonality
Breaking (DSOB) technique proposed in this Letter is in essence
insensitive to birefringence and polarization rotation, thus enabling
remote sensing through optical fibers, in which stress/torsion-induced
birefringences are usually highly detrimental to usual polarimetric
measurements \cite{des09}.

\begin{figure}[htbp]
\includegraphics[width=8cm]{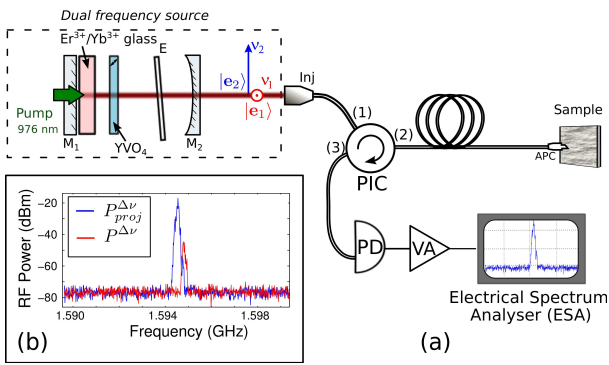}
\caption{(a) Schematic representation of the depolarization
  remote-sensing setup. Dashed box: description of the Er,Yr;Glass
  laser producing the two orthogonally polarized and frequency shifted
  fields. (b) Measurement of the RF contrast obtained on a fibred
  mirror ($\Delta \nu=1.254$ GHz). \label{montage}}
\end{figure}

In order to experimentally illustrate and validate the DSOB principle,
we have implemented the setup depicted in Fig.\ref{montage}.(a). In
this setup, the probe field state is prepared using a single laser
source which inherently produces the two orthogonal polarization
states with shifted frequencies. This laser consists of an Er,Yb;Glass
4-cm-long external cavity laser emitting at $1550$ nm. Single
longitudinal mode oscillation is obtained with an intracavity
$40$-$\mu$m-thick silica etalon whose both sides are coated for $40\%$
reflection at $1550$ nm. Moreover, a $500$-$\mu$m-thick YVO$_4$
crystal, cut at $45^{\mathrm{o}}$ of its optical axis, is inserted
into the laser cavity in order to, at the same time, define two linear
polarization eigenstates, lift the frequency degeneracy, and finally
ensure a slight polarization walk-off ($50$ $\mu$m) in the active
medium. This walk-off reduces the nonlinear coupling between the two
eigenstates \cite{bai09} leading to robust and simultaneous
oscillation of the two polarization eigenstates. The active medium is
diode pumped at $976$ nm. Lateral positioning of the pump beam enables
equalization between the two eigenstates intensities. By slightly
tilting the etalon and the YVO$_4$ crystal, it is possible to set the
frequency difference $\Delta\nu = \nu_2- \nu_1$ to a value compatible
with the detection setup, namely within the radio-frequency (RF)
range, i.e., $\Delta \nu< 2$ GHz. With this configuration, the laser
provides an output power of $1.8$ mW with a pump power of
approximately $130$ mW.

Let us remind that the depolarization measurement concept we propose
here is expected to be well suited for remote sensing. In order to
validate this expectation, the laser output is injected into, and
guided along, a 2 m-long single-mode optical fiber (SMF28) before
shining the sample. The backscattered light is back-propagated into
the same fiber, extracted with a polarization-insensitive circulator
(PIC) and directed on a high-band pass ($16$ GHz) photodiode (PD). The
detected RF signal is amplified with a high-gain (60 dB, 2 GHz
cut-off frequency) voltage amplifier and analyzed on a 40 GHz band
pass electrical spectrum analyzer (ESA). It is worth noticing that
this measurement setup does not require any component to be inserted
at the distal fiber-end and may thus be directly adapted to commercial
endoscopes. More importantly, the setup can be operated at any
wavelength since no polarizing or birefringent elements are
needed. The spectral range is thus only limited by the source, the
photodiode and the fiber spectral excursions.

To calibrate and test this experimental setup, we first evaluated the
maximum ``polarimetric orthogonality contrast'' available with such a
measurement scheme. This was done by comparing the residual beat note
power $P^{\Delta\nu}$ (caused by imperfect orthogonality between the
polarization states) to the maximum beat note power available
$P_{proj}^{\Delta\nu}$, obtained by inserting a polarizer
(transmission $T_{p}=85 \%$) at a $45^\mathrm{o}$ angle with respect
to the illumination polarization directions at the laser output. This
contrast was derived from the corresponding spectra analyzed around
frequency $\Delta \nu$ on the ESA, as illustrated in
Fig.\ref{montage}.(b). After injection into the fiber, we measured a
high contrast of $-34\pm1$ dB. After propagation through a 20 km-long
SMF, a reduced but still high contrast of $-25\pm1$ dB was measured,
showing that orthogonality is fairly maintained during propagation
over tens of kilometers. It was also observed that the circulator was
rather detrimental to orthogonality, since a contrast of $-28\pm1$ dB
has been measured at the output port (2) of this component, and
$-27\pm1$ dB at the output port (3) after reflection on a fibred
mirror. At this level, it is interesting to note that such setup
providing a measurement dynamics of $25$ dB makes it possible to measure
values of $\degp$ up to $99.5\, \%$ (respectively $99.95\,\%$ with a
$30$ dB dynamics). The new DSOB technique addressed in this Letter may
therefore be a very efficient tool when slight depolarization
contrasts have to be characterized.

\begin{figure}[htbp]
\includegraphics[width=5cm]{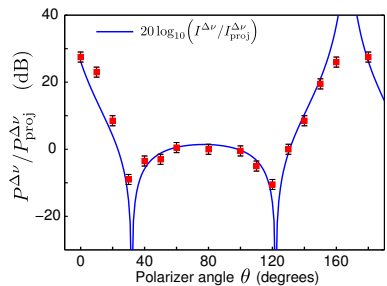}
\caption{Polarimetric orthogonality contrast measured with the setup
  of Fig.\ref{montage}.(a) on a rotating polarizer and mirror. The RF
  powers were measured with an ESA. \label{fig3}}
\end{figure}

Further calibration was also conducted by using the setup of
Fig.\ref{montage}.(a) and a fibred collimator to illuminate a
rotating polarizer (axis $\theta_0=32^\mathrm{o}$) followed by a
mirror.  The polarimetric contrast of the light back-propagated into
the fiber was analyzed on the spectrum analyzer for various orientations
$\theta$ of the polarizer. The obtained results are plotted in Fig. \ref{fig3} and
are in fair agreement with our theoretical predictions showing that in the
experimental configuration considered, one has
\begin{equation}
\frac{I^{\Delta\nu}}{I^{\Delta\nu}_{\mathrm{proj}}}=\Biggl|\frac{2\Bigl[\sin\bigl(2(\theta-\theta_0)\bigr)+K_0\Bigr]}{T_{p}\Bigl[1+K_0+\sin\bigl(2(\theta-\theta_0)\bigr)\Bigr]}\Biggr|
\end{equation}
with $K_0=0.3 \,\%$ denoting the amount of intensity reflected at the
fiber/collimator end.

\begin{figure}[htbp]
\includegraphics[width=7cm]{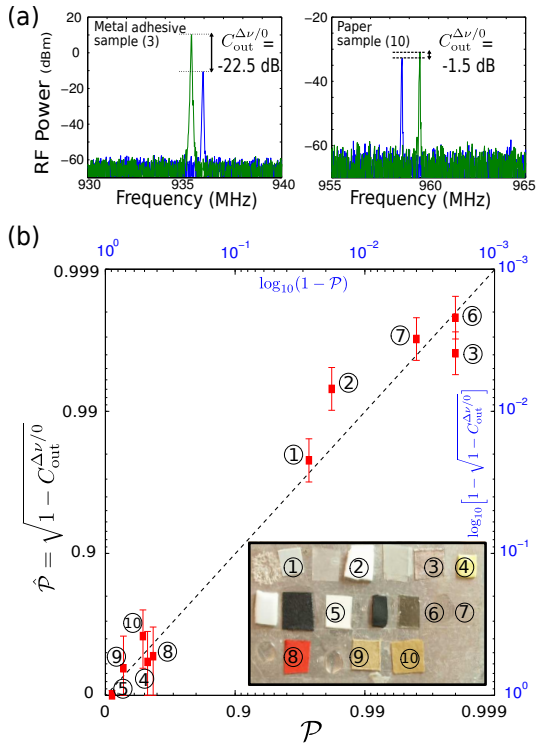}
\caption{(a) Evaluation of the RF polarimetric contrast on a non
  depolarizing (left) and a depolarizing (right) sample with $\Delta
  \nu \simeq 950$ MHz. (b) The DOP estimated ($\hat{\degp}$) from RF beat note
  power measurement is in fair agreement with the DOP of the samples
  characterized at $\lambda= 1.55 \ \mu$m with a standard free-space Stokes
  measurement setup \label{fig4}}
\end{figure}

Once calibrated, the setup of Fig.\ref{montage}.(a) was tested on a
variety of materials (see inset of Fig.\ref{fig4}.(b)) to validate
experimentally the DSOB technique. To match endoscopic applications
requirements, we considered a challenging experimental configuration
where the materials under investigation were placed in the vicinity of
the APC fiber-end connector without any collimation optics. In such
conditions, a very weak proportion of the illumination optical power
was backscattered into the fiber mode and propagated to the detector:
$\simeq 10\ \mu$W on a metallic adhesive sample (sample 3) and only
$\simeq 0.7 \ \mu$W on a diffusive paper sheet (sample 5). However,
the heterodyne detection scheme involved in DSOB enables a very high
sensitivity after amplification of the detected signal and analysis on
the ESA. A natural improvement of this setup would be using a lock-in
detection to measure precisely the output electrical power at
frequency $\Delta\nu$. Indeed, the ESA used here for the proof of
concept is not required and could be advantageously replaced by a
demodulation circuit since the frequency of interest $\Delta\nu$ is
known.


The results obtained are reported in Fig.\ref{fig4}. It can first be
checked in Fig.\ref{fig4}.(a) that the relative beat note optical
power is low ($C_\mathrm{out}^{\Delta \nu/0}=-22.5 \pm 1$ dB) on
a non-depolarizing medium (metal adhesive, $\degp=0.99$), whereas it
considerably increases ($C_\mathrm{out}^{\Delta \nu/0}=-1.5 \pm
1$ dB) on a diffusive and depolarizing sample (white paper,
$\degp=0.11$).  For the sake of experimental convenience,
$C_\mathrm{out}^{\Delta \nu/0}$ was determined here by
normalizing the RF output power $P_\mathrm{out}^{\Delta \nu}$ (blue
curves) by $P_\mathrm{out,proj}^{\Delta \nu}$ (green curves), i.e.,
the RF power obtained with projected polarization states at the laser
output, which provides the overall optical power backscattered by the
sample (up to a factor $T_{p}/2$).  Then, on a variety of ten samples
with distinct depolarization properties, the relative beat note
optical power $\mathrm{p}_\mathrm{out}^{\Delta \nu}$ was measured and
injected in Eq.(\ref{defC}) to provide an estimation of the DOP. The
estimated values of $\hat{\degp}$ are plotted in Fig.\ref{fig4}.(b)
and are in good agreement with the DOP of the samples, characterized
independently with standard Stokes free-space measurements performed
at $1.55\ \mu$m. These experimental results validate the concept of
remote depolarization sensing by polarization orthogonality breaking,
as well as the DSOB setup proposed.

In this Letter, we introduced a new polarimetric measurement
technique, based on the concept of orthogonality breaking, and
allowing one to measure the depolarization strength of a material from
a single measurement in a few tens of milliseconds. In addition, the
DSOB technique is \emph{per se} insensitive to propagation through a
fiber and is easily implemented without requiring any specific
component at the distal end of the fiber, nor in front of the detector,
thus paving the way for depolarization remote-sensing of biological
tissues with conventional endoscopes. Based on a heterodyne detection
setup, the proposed method is highly sensitive and therefore perfectly
suited for biomedical applications where biological tissues are often
slightly depolarizing and prone to photodamage. Improving the source
control and implementing a demodulation circuit is a perspective to
this work, before applying the DSOB technique to real-time
polarization-sensitive endoscopic imaging on samples of biological
interest.

%

\end{document}